# Dielectric Evidence for Possible Type-II Multiferroicity in α-RuCl$_3$

Jia Cheng Zheng,[1, 2] Yi Cui,[2] Tian Run Li,[2] Ke Jing Ran,[3] Jinsheng Wen,[3,4]

and Weiqiang Yu[2,*]

[1]*Department of Physics, Beijing Jiaotong University, Beijing 100044, China*
[2] *Department of Physics and Beijing Key Laboratory of Opto-electronic Functional Materials & Micro-nano Devices, Renmin University, Beijing 100872, China*
[3]*National Laboratory of Solid State Microstructures and Department of Physics, Nanjing University, Nanjing 210093, China*
[4]*Innovative Center for Advanced Microstructures, Nanjing University, Nanjing 210093, China*

The Kitaev model in a honeycomb lattice, which has an exactly solvable spin liquid as the ground state [1], has attracted a lot of research interests. Recently, Kitaev-type interactions were discovered in several quasi-2D, honeycomb lattice compounds, such as $A_2$IrO$_3$ ($A$=Li, Na) [2–7] and α-RuCl$_3$ [8–15]. In these compounds, both the spin-orbit coupling and electron correlations play an essential role in the emergent Mott insulator behaviors [2,16–19]. Although the ground states of these compounds are magnetically ordered [20–25] due to non-Kitaev interactions, proximate Kitaev spin liquid behavior was proposed at high energies [23–26], whereas the low-energy spin fluctuations are strongly affected by the interplay of non-Kitaev terms [27,28].

For α-RuCl$_3$, the Kitaev terms are found to be very strong, comparing to the non-Kitaev terms including the Heisenberg exchange couplings and off-diagonal exchange couplings [15,29,30]. In particular, the magnetic ordering can be easily suppressed by an external magnetic field, giving rise to a quantum phase transition at a critical field $H_c \approx 7.6$ T [27,28,31] with field applied in the ab-plane of the lattice, or much higher fields applied out of the plane [32]. Recently, it has reported that the magnetic ordering can also be completely suppressed with a pressure above 1 GPa [33,34], leading to a novel magnetically disordered phase [34]. All these studies indicate a strong coupling between lattice, spin, and orbital degrees of freedom.

This paper presents dielectric constant measurements on α-RuCl$_3$ single crystals. An anomalous reduction was found in the dielectric constant ε when the system enters the magnetic ordering upon cooling. When the magnetic ordering is suppressed by the magnetic field, the reduction in $\varepsilon$ is absent. Simultaneously, $\varepsilon$ also shows an anomalous reduction when the system undergoes the structural transition. Our data reveal a strong coupling among the charge and the magnetism of the system, and the dielectric constant can be used to probe the magnetic ordering and the quantum phase transition of this honeycomb lattice antiferromagnet.

The single crystal was grown by the chemical vapor transport method [15]. The sample is plate like, with the crystalline c-axis along the thinnest dimension. The high quality of sample is demonstrated by the neutron scattering [15] and the NMR [27] study on the samples grown by the same group. For dielectric measurements at the ambient pressure, a single crystal with a dimension of 18mm*8mm*0.2mm is chosen.

The dielectric measurements are performed by a capacitance method. Two copper plates were respectively attached to the two cleavage surfaces (along the ab-plane) of the single crystal, and the capacitance between two plates was measured with an Agilent 4263B LCR meter with an excitation level of 1.0 V at 100 kHz. The dielectric constant was then calculated by $\varepsilon = Cd / S$, where $C$ represents the capacitance measured between two copper plates, $d$ represents the distance between copper plates, and $S$ represents the effective surface area of one copper plate. In this paper, the relative value $\varepsilon / \varepsilon_0$ are presented for all figures, where $\varepsilon_0$ is the dielectric constant of vacuum. The d and S were measured at the ambient conditions, whose changes under field and temperature affect were not considered in the calculation of $\varepsilon$.

The dielectric constant was first measured at the ambient field. In Fig. 1, the ac dielectric constant $\varepsilon$ along the crystalline c-axis is shown as a function of temperature from 200 K down to 2 K. Upon cooling, $\varepsilon$ decreases monotonically. By a first derivation of the data, as shown in the inset of Fig. 1, two peaks are shown at temperature about 170 K and 7.5 K. The high-temperature peak is consistent with a structural

*Corresponding author (email: wqyu_phy@ruc.edu.cn)

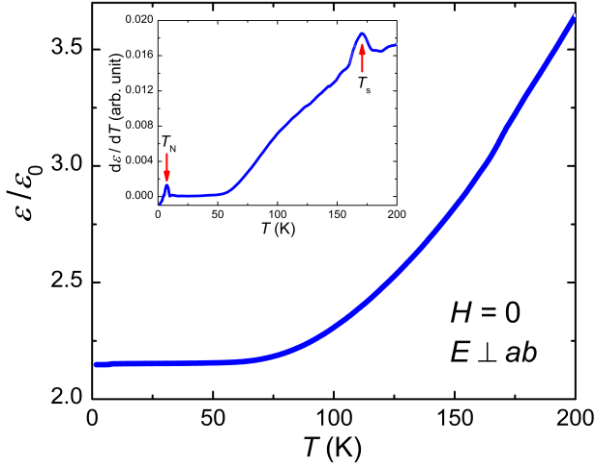

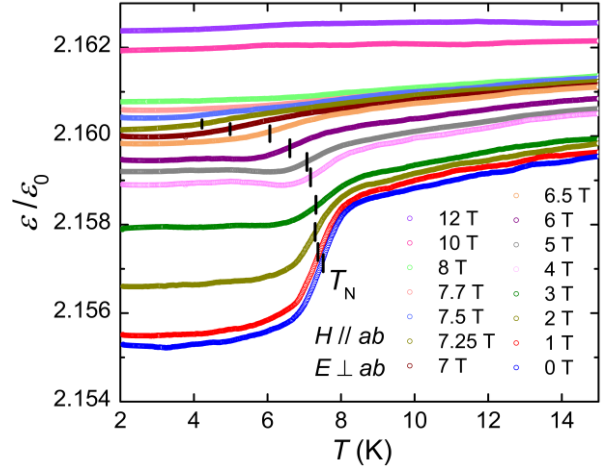

**Figure 1** The ac dielectric constant $\varepsilon$ of α-RuCl$_3$ single crystal measured under zero magnetic field. $\varepsilon_0$ is the dielectric constant of vacuum. The electric field is applied along the c-axis of the crystal. Inset: The first derivation of $\varepsilon/\varepsilon_0$. The down arrow points at the magnetic transition at $T_N$(~7.5 K) and the up arrow points at the structural transition at $T_s$(~170 K).

**Figure 2** The dielectric constant $\varepsilon$ of the sample measured under different magnetic fields, applied in the ab-plane of the crystal. The black vertical lines mark the anomaly decrease in $\varepsilon$, demonstrating the magnetic transitions under different fields.

transition with changing stacking pattern along the c-axis, as revealed by earlier reports in α-RuCl$_3$ [12,35]. When cooled to about 7.5 K, the $d\varepsilon/dT$ exhibits another peak, as shown in the inset of Fig. 1. Earlier magnetization and neutron scattering studies have revealed magnetic ordering of α-RuCl$_3$ crystals, whose transition temperature $T_N$ depends on the stacking pattern along the c-axis [12,23,25,29]. The $T_N$ is about 14 K for the AB stacking and 7.5 K for the ABC stacking. Our sample is primarily composed of the ABC stacking [15,27]. Therefore, the coincidence of the peak of $\varepsilon$ at 7.5 K is caused by the magnetic transition of the sample with the ABC stacking. The microscopic origin for this dielectric anomaly at the magnetic transition suggests a coupling between magnetic and charge properties, which is extensively discussed later.

We further verify that the above anomaly is always seen in the magnetic transitions of α-RuCl$_3$, by applying an external magnetic field. In Fig. 2, the dielectric constant is shown as functions of temperature, with different magnetic fields applied in the ab-plane of the sample. From 15 down to 2 K, $\varepsilon$ demonstrates an anomalous sharp decrease with temperature for all fields up to 7.5 T. The onset temperature for this anomaly, as marked by the black vertical lines, decreases with field. In fact, earlier NMR and the specific heat studies have shown that the magnetic ordering diminishes when the external field exceeds a critical field $H_c \approx 7.6$ T [27]. For all fields, our onset temperatures for the decreases of $\varepsilon$ coincide with the $T_N$. Therefore, our data confirm that the sharp drop of $\varepsilon$ always accompanies the magnetic transition.

The relation between the dielectric anomaly and the magnetic order was further investigated by the field dependence of $\varepsilon$ at different temperatures. In Fig. 3, the $\varepsilon$ values are shown as functions of fields. At 2 K, a two-stage drop of $\varepsilon$ is seen with a decreasing field. The $\varepsilon$ values show a high-field decrease at about 7.5 T for all temperatures up to 7 K, which is consistent with the critical field $H_c$ to suppress the magnetic order, as reported by NMR [27]. With increasing $T$, the fields at the drop of $\varepsilon$, as indicated by black lines, are consistent with the critical field at each temperature again by earlier reports [27,28,36]. A second drop of $\varepsilon$ appears at lower fields, as marked by the pink vertical lines in Fig. 3. This second drop occurs at a nearly constant field at different temperatures.

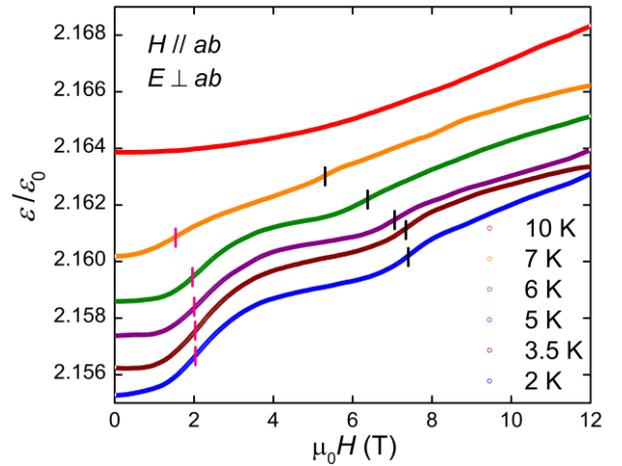

**Figure 3** Magnetic field dependence of the dielectric constant measured at fixed temperatures. The black vertical lines mark the fields below which the sample orders magnetically. A second drop of $\varepsilon$ at a lower field at each temperature is indicated by the pink vertical lines.

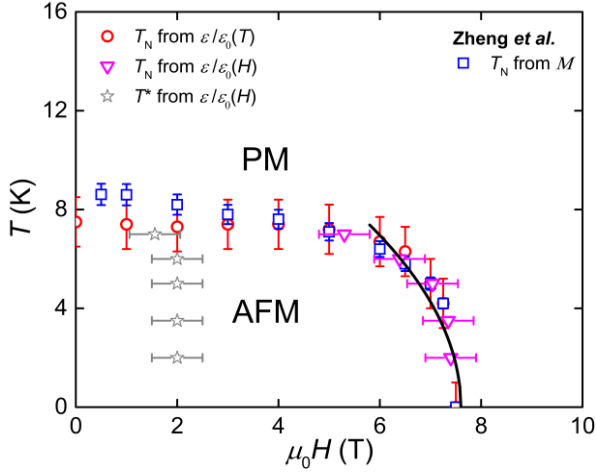

**Figure 4** Magnetic phase diagram of α-RuCl$_3$ determined by ε(T,H). An earlier phase diagram by magnetization measurement [27] are also plotted for comparison. The solid line is mean-field function fit to $T_N(H)$ data. $T_N$ are determined from dielectric constant with field applied in the *ab* plane shown in Fig. 2 and 3.

The phase diagram of α-RuCl$_3$ determined by our dielectric data is shown in Fig. 4. These data are also consistent with earlier magnetization measurements [27], as shown in Fig. 4. Therefore, our data clearly demonstrate that the anomaly in the dielectric constant is caused by the magnetic ordering in α-RuCl$_3$, and ε can be a simple probe for magnetic transitions at the ambient condition and quantum phase transitions under external field. The $T_N(H)$ follows a mean-field function $T_N \sim (H_c - H)^{1/2}$ which supports a second order phase transition. Therefore, a quantum critical point is strongly suggested at $H_c$ by our study.

$T^*(H)$, defined as the onset temperature, was then plotted for a second drop of ε(H) under field at different temperatures. Since this occurs below $T_N$, this anomaly may be caused by some inhomogeneity of the system or stacking faults of the crystal. Local measurements, such as micro force microscopy(MFM), are requested to understand the origin of this low temperature anomaly in ε.

In principle, the dielectric constant is affected by the ionic position, the electronic conductivity, and the dimension of the sample, all of which change with temperature. To understand the anomalous behavior of ε close to the magnetic transition in α-RuCl$_3$, all possible contributions to our ε data are discussed below.

i) A structural effect. Although a structural transition with changing stacking pattern along the *c*-axis has been reported at about 150 K in α-RuCl$_3$ [12,35], our ε data only detects a very small anomaly in dε / dT in this temperature range. Furthermore, upon cooling, the lattice parameter *c* should in principle shrink with a larger extent than *a* or *b* for this quasi-2D materials because of a van der Waals coupling along the *c*-axis [24,37]. As a result, an increase of measured capacitance is expected. However, such a behavior is not seen in our measured dielectric constant. Therefore, the measured dielectric constant is not strongly affected by the change of lattice parameters and the large kinked drop of ε close to the magnetic transition cannot be attributed to the change of lattice parameters, especially when the magnetic transition of α-RuCl$_3$ is a second-order type at zero-pressure, where the change of the lattice parameters ought to be very small.

ii) Weak electronic conductivity. For α-RuCl$_3$, a Mott gap is about 1 eV [14,38], and thermal activated conductivity is found by the high-pressure transport measurements [33]. In such cases, the finite conductivity should strongly enhance the measured ε and lead to a decrease of ε upon cooling for a gapped system because of reduced thermal activations. This contribution seems to qualitatively fit in our high-temperature data: as shown in Fig. 1, the measured ε at 180 K is over 10 times of the vacuum; upon cooling, ε is rapidly reduced when the temperature is decreased to 50 K.

Below 50 K, however, ε levels off with a large value, which suggests that an additional contribution, other than conductive electrons, is effective.

iii) Magnetoelectric coupling. This is in analogy to the type-II multiferroics, whose ferroelectricity is driven by magnetic ordering just below the magnetic ordering temperature [39]. However, in such cases, spiral magnetism or other types of non-collinear magnetic structures are requested to produce the charge polarization [39]. For α-RuCl$_3$, a collinear zigzag magnetic ordering is established in each layer [25,29], which is different from most type-II multiferroics. However, this study shows a dramatic drop of the dielectric constant just at the magnetic transition, analogous to other type-II multiferroics. Hence, α-RuCl$_3$ is also a likely type-II multiferroics with magnetic driven charge polarization. Indeed, the drop of the ε below $T_N$ in α-RuCl$_3$ suggests an antiferroelectric phase rather than a ferroelectric type, as seen in Cu$_3$Bi(SeO$_3$)$_2$O$_2$Cl [40].

Recently, an anisotropic magnetodielectric measurement has also been reported in α-RuCl$_3$ [41]. By contrast, their drop of ε is consistent with the AB stacking with a higher $T_N$ (~ 14 K). Our crystals are confirmed to be of ABC stacking from previous studies and therefore the same $T_N$ (~ 7.5 K) by different probes drawing a direct coupling between magnetism and dielectric properties.

In addition, the interlay coupling of α-RuCl$_3$ is a Van der Waals type [25], which in principle is too weak to account for such a strong magnetodielectric coupling. Aoyama *et al.* [41] proposed that the antiferroelectricity in α-RuCl$_3$ originates from the in-plane zigzag magnetic ordering. The competition between the broken inversion symmetry among the neighboring zigzag magnetic chains and the remaining inversion symmetry, causes an anti-parallel ionic displacement among the neighboring Ru$^{3+}$ sites along each zigzag chain. As a result, the local polarizations of the nearest-neighbor $Ru^{3+}$ ions are antiparallel.

In summary, we have resolved a sudden change of the dielectric constant in α-$RuCl_3$ at the magnetic ordering temperature and below a critical magnetic field which suppresses the magnetic ordering. This change is not seen at high fields, when the magnetic transition is absent. Our data also suggests a second-order quantum phase transition. Although an exact understanding for this novel observation requests further studies, this work suggests that α-$RuCl_3$ is a possible type-II antiferroelectric material and reveals dielectric measurement as an alternative probe for magnetic phase transition in α-$RuCl_3$.

*This Work is supported by the Ministry of Science and Technology of China (Grant No. 2016YFA0300504), the National Science Foundation of China (Grant No.11374364) and the Fundamental Research Funds for the Central Universities and the Research Funds of Renmin University of China (Grant No. 14XNLF08).*

1. A. Kitaev, Ann. Phys. **321**, 2 (2006).
2. G. Jackeli, and G. Khaliullin, Phys. Rev. Lett. **102**, 017205 (2009).
3. J. Chaloupka, G. Jackeli, and G. Khaliullin, Phys. Rev. Lett. **105**, 027204 (2010).
4. Y. Singh, S. Manni, J. Reuther, T. Berlijn, R. Thomale, W. Ku, S. Trebst, and P. Gegenwart, Phys. Rev. Lett. **108**, 127203 (2012).
5. H. Gretarsson, J. P. Clancy, X. Liu, J. P. Hill, E. Bozin, Y. Singh, S. Manni, P. Gegenwart, J. Kim, A. H. Said, D. Casa, T. Gog, M. H. Upton, H.-S. Kim, J. Yu, V. M. Katukuri, L. Hozoi, J. van den Brink, and Y.-J. Kim, Phys. Rev. Lett. **110**, 076402 (2013).
6. H. S. Chun, J.-W. Kim, J. Kim, H. Zheng, C. C. Stoumpos, C. D. Malliakas, J. F. Mitchell, K. Mehlawat, Y. Singh, Y. Choi, T. Gog, A. Al-Zein, M. M. Sala, M. Krisch, J. Chaloupka, G. Jackeli, G. Khaliullin, and B. J. Kim, Nature Phys. **11**, 462 (2015).
7. J. G. Rau, E. K.-H. Lee, and H.-Y. Kee, Ann. Rev. Condens. Matter Phys. **7**, 195 (2016).
8. I. Pollini, Phys. Rev. B **53**, 12769 (1996).
9. K. W. Plumb, J. P. Clancy, L. J. Sandilands, V. V. Shankar, Y. F. Hu, K. S. Burch, H.-Y. Kee, and Y.-J. Kim, Phys. Rev. B **90**, 041112 (2014).
10. H.-S. Kim, V. Shankar, A. Catuneanu, and H.-Y. Kee, Phys. Rev. B **91**, 241110 (2015).
11. L. J. Sandilands, Y. Tian, K. W. Plumb, Y.-J. Kim, and K. S. Burch, Phys. Rev. Lett. **114**, 147201 (2015).
12. Y. Kubota, H. Tanaka, T. Ono, Y. Narumi, and K. Kindo, Phys. Rev. B **91**, 094422 (2015).
13. L. J. Sandilands, Y. Tian, A. A. Reijnders, H.-S. Kim, K. W. Plumb, Y.-J. Kim, H.-Y. Kee, and K. S. Burch, Phys. Rev. B **93**, 075144 (2016).
14. A. Koitzsch, C. Habenicht, E. M•uller, M. Knupfer, B. B•uchner, H. C. Kandpal, J. van den Brink, D. Nowak, A. Isaeva, and T. Doert, Phys. Rev. Lett. **117**, 126403 (2016).
15. K. J. Ran, J. H. Wang, W. Wang, Z.-Y. Dong, X. Ren, S. Bao, S. C. Li, Z. Ma, Y. Gan, Y. T. Zhang, J. T. Park, G. C. Deng, S. Danilkin, S.-L. Yu, J.-X. Li, and J. S. Wen, Phys. Rev. Lett **118**, 107203 (2017).
16. B. J. Kim, H. Ohsumi, T. Komesu, S. Sakai, T. Morita, H. Takagi, and T. Arima, Science **323**, 13295 (2009).
17. K. Foyevtsova, H. O. Jeschke, I. I. Mazin, D. I. Khomskii, and R. Valentí, Phys. Rev. B **88**, 035107 (2013).
18. T. Birol, and K. Haule, Phys. Rev. Lett. **114**, 096403 (2015).
19. J. Chaloupka, and G. Khaliullin, Phys. Rev. B **94**, 064435 (2016).
20. J. M. Fletcher, W. E. Gardner, A. C. Fox, and G. Topping, J. Chem. Soc. A 1038 (1967).
21. F. Ye, S. X. Chi, H. B. Cao, B. C. Chakoumakos, J. A. Fernandez-Baca, R. Custelcean, T. F. Qi, O. B. Korneta, and G. Cao, Phys. Rev. B **85**, 180403 (2012).
22. S. K. Choi, R. Coldea, A. N. Kolmogorov, T. Lancaster, I. I. Mazin, S. J. Blundell, P. G. Radaelli, Y. Singh, P. Gegenwart, K. R. Choi, S.-W. Cheong, P. J. Baker, C. Stock, and J. Taylor, Phys. Rev. Lett. **108**, 127204 (2012).
23. J. A. Sears, M. Songvilay, K. W. Plumb, J. P. Clancy, Y. Qiu, Y. Zhao, D. Parshall, and Y.-J. Kim, Phys. Rev. B **91**, 144420 (2015).
24. R. D. Johnson, S. C. Williams, A. A. Haghighirad, J. Singleton, V. Zapf, P. Manuel, I. I. Mazin, Y. Li, H. O. Jeschke, R. Valentí, and R. Coldea, Phys. Rev. B **92**, 235119 (2015).
25. H. B. Cao, A. Banerjee, J.-Q. Yan, C. A. Bridges, M. D. Lumsden, D. G. Mandrus, D. A. Tennant, B. C. Chakoumakos, and S. E. Nagler, Phys. Rev. B **93**, 134423 (2016).
26. M. Majumder, M. Schmidt, H. Rosner, A. A. Tsirlin, H. Yasuoka, and M. Baenitz, Phys. Rev. B **91**, 180401 (R) (2015).
27. J. C. Zheng, K .J. Ran, T. R. Li, J. H. Wang, P. S. Wang, B. Liu, Z. X. Liu, B. Normand, J. S. Wen, and W. Q. Yu, Phys. Rev. Lett. **119**, 227208 (2017).
28. I. A. Leahy, C. A. Pocs, P. E. Siegfried, D. Graf, S.-H. Do, K.-Y. Choi, B. Normand, and M. Lee, Phys. Rev. Lett. **118**, 187203 (2017).
29. A. Banerjee, C. A. Bridges, J. Q. Yan, A. A. Aczel, L. Li, M. B. Stone, G. E. Granroth, M. D. Lumsden, Y. Yiu, J. Knolle, S. Bhattacharjee, D. L. Kovrizhin, R. Moessner, D. A. Tennant, D. G. Mandrus, and S. E. Nagler, Nature Mater. **15**, 733 (2016).
30. A. Banerjee, J. Q. Yan, J. Knolle, C. A. Bridges, M. B. Stone, M. D. Lumsden, D. G. Mandrus, D. A. Tennant, R. Moessner, and S. E. Nagler, Science **356**, 1055 (2017).
31. R. Hentrich, A. U. B. Wolter, X. Zotos, W. Brenig, D. Nowak, A. Isaeva, T. Doert, A. Banerjee, P. Lampen-Kelley, D. G. Mandrus, S. E. Nagler, J. Sears, Y.-J. Kim, B. Büchner, and C. Hess, arXiv:1703.08623 (2017).
32. S.-H. Baek, S.-H. Do, K.-Y. Choi, Y. S. Kwon, A. U. B. Wolter, S. Nishimoto, J. van den Brink, and B. Büchner, Phys. Rev. Lett. **119**, 037201 (2017).
33. Z. Wang, J. Guo, F. F. Tafti, A. Hegg, S. Sen, V. A. Sidorov, L. Wang, S. Cai, W. Yi, Y. Zhou, H. Wang, S. Zhang, K. Yang, A. Li, X. Li, Y. Li, J. Liu, Y. Shi, W. Ku, Q. Wu, R. J. Cava, L. Sun, arXiv: 1705.06139 (2017).
34. Y. Cui, J. Zheng, K. Ran, J. S. Wen, Z.-X. Liu, B. Liu, W. A. Guo, and W. Q. Yu, Phys. Rev. B **96**, 205147 (2017).
35. M. Ziatdinov, A. Banerjee, A. Maksov, T. Berlijn, W. Zhou, H. B. Cao, J.-Q. Yan, C. A. Bridges, D. G. Mandrus, S. E. Nagler, A. P. Baddorf, and S. V. Kalinin, Nat. Commun. **7**, 13774 (2016).
36. J. A. Sears, Y. Zhao, Z. Xu, J. W. Lynn, and Y.-J. Kim, Phys. Rev. B **95**, 180411 (2017).
37. H.-S. Kim, and H.-Y. Kee, Phys. Rev. B **93**, 155143 (2016).
38. X. Q. Zhou, H. X. Li, J. A. Waugh, S. Parham, H.-S. Kim, J. A. Sears, A. Gomes, H.-Y. Kee, Y.-J. Kim, and D. S. Dessau, Phys. Rev. B **94**, 161106 (2016).
39. S.-W. Cheong, and M. Mostovoy, Nature Mater. **6**, 13 (2007).
40. H. C. Wu, K. Devi Chandrasekhar, J. K. Yuan, J. R. Huang, J.-Y. Lin, H. Berger, and H. D. Yang, Phys. Rev. B **95**, 125121 (2017).
41. T. Aoyama, Y. Hasegawa, S. Kimura, T. Kimura, and K. Ohgushi, Phys. Rev. B **95**, 245104 (2017).